\documentclass[%
 reprint,
 superscriptaddress,
 amsmath,amssymb,
 aps,
 prl
]{revtex4-2}

\usepackage{graphicx}
\usepackage{dcolumn}
\usepackage{bm}

\usepackage{siunitx}

\begin{document}

\preprint{APS/123-QED}

\title{Nonadiabatic Molecular Dynamics on Real-time  \\ Excited-State Surfaces
via Machine Learning Hamiltonians}

\author{Changwei Zhang}
\author{Yang Zhong}
\author{Zhi-Guo Tao}
\affiliation{Key Laboratory of Computational Physical Sciences (Ministry of Education), Institute of Computational Physical Sciences, State Key Laboratory of Surface Physics, and Department of Physics, Fudan University, Shanghai, 200433, China.}

\author{Yingzhou Li}
\affiliation{School of Mathematical Sciences, Fudan University, Shanghai, 200433, China}
\author{Zhenggang Lan}
\affiliation{SCNU Environmental Research Institute, Guangdong Provincial Key Laboratory of Chemical Pollution and Environmental Safety \& MOE Key Laboratory of Environmental Theoretical 
Chemistry, South China Normal University, Guangzhou, Guangdong, 510006, China.}

\author{Oleg V. Prezhdo}
\affiliation{Department of Chemistry and Chemical Biology, University of New Mexico, Albuquerque, New Mexico, 87106, United States.}

\author{Xin-Gao Gong}
\author{Weibin Chu}
\email{wbchu@fudan.edu.cn}
\author{Hongjun Xiang}
\email{hxiang@fudan.edu.cn}
\affiliation{Key Laboratory of Computational Physical Sciences (Ministry of Education), Institute of Computational Physical Sciences, State Key Laboratory of Surface Physics, and Department of Physics, Fudan University, Shanghai, 200433, China.}

\date{\today}

\begin{abstract}
Simulating the coupled, nonequilibrium dynamics of electrons and nuclei is a central challenge in chemistry, physics, and materials science, governing phenomena from photocatalysis to quantum information. The primary bottleneck has been the lack of a general, accurate, and efficient method for modeling the complete excited-state landscape: the potential energy surfaces, forces, and non-adiabatic couplings for multiple electronic states. While machine learning has revolutionized ground-state simulations and shown promise for excited states in molecules, a unified framework that solves the complete multi-state problem for general condensed matter systems has remained elusive. Here we introduce on-the-fly N\textsuperscript{2}AMD (Neural network NAMD), a machine learning framework that makes on-the-fly NAMD in solids a reality. By employing an equivariant neural network to predict the system Hamiltonian, the framework delivers excited-state energies, forces, and non-adiabatic coupling vectors at a fraction of the cost of ab initio calculations. Crucially, it allows simulations with hybrid functional accuracy, a level of approach previously inaccessible for NAMD. We showcase its capabilities with three topical examples: correcting order-of-magnitude errors in carrier dynamics predicted by conventional procedure in a MoS$_2$/WS$_2$ heterostructure, simulating previously inaccessible photoinduced ferroelectric switching, and capturing real-time polaron formation in TiO$_2$ at the hybrid-functional level. On-the-fly N\textsuperscript{2}AMD moves beyond the limitations of equilibrium theory, establishing a new paradigm for the predictive, first-principles design of materials operating far from equilibrium.
\end{abstract}

\maketitle

The coupled, non-equilibrium dynamics of electrons and nuclei constitute the microscopic foundation of modern condensed matter physics, governing phenomena ranging from photovoltaic energy conversion and photocatalysis to ultrafast optoelectronics \cite{Williams_solid-state_1984, volkov_attosecond_2019, akimov_theoretical_2013, jhalani_ultrafast_2017, long_nonadiabatic_2017, guan_theoretical_2022,wang_surface_2020}. However, developing predictive, first-principles theoretical frameworks capable of describing materials driven far from equilibrium remains a formidable challenge. Specifically, intense photoexcitation invalidates standard perturbative approaches, necessitating methods that treat electronic and lattice degrees of freedom on an equal footing.

Non-adiabatic molecular dynamics (NAMD), particularly via the fewest-switches surface hopping (FSSH) algorithm \cite{tully_molecular_1990}, has emerged as the fundamental tool for simulating such dynamics in molecules and small nanoclusters \cite{nelson_non-adiabatic_2020, Polyak_ultrafast_2019, mai_Nonadiabatic_2018}. However, extending rigorous FSSH to solids presents a significant computational challenge, as exact "on-the-fly" propagation requires calculating excited-state forces and non-adiabatic coupling vectors (NACVs) at every time step across thousands of trajectories. Consequently, solid-state NAMD has relied almost exclusively on the Classical Path Approximation (CPA), also known as the Neglect of Back-Reaction Approximation (NBRA) \cite{akimov_pyxaid_2013, smith_modeling_2019}. 
By decoupling nuclear motion from the instantaneous electronic state and treating phonons as a precomputed background, the CPA assumes that the lattice evolves exclusively on a static potential energy surface (PES), therefore replacing computationally demanding NACVs with scalar couplings along a fixed trajectory, and bypassing the need for excited-state forces.
While extensively employed within the solid-state community for modeling weak-field excitations, the CPA suffers from a critical physical deficiency: it entirely neglects the back-reaction of excited carriers on the lattice, rendering it blind to the profound reshaping of interatomic forces under electronic excitation. 
As a result, the CPA collapses in the far-from-equilibrium regimes central to NAMD, failing to capture phenomena where the lattice response to the excited electron is the primary driver, such as photo-induced phase transitions, chemical reactions, ferroelectric switching, and the dynamic formation of quasiparticles (such as polarons, polaritons and magnons). Overcoming this limitation requires restoring the essential Tully's feedback loop by explicitly evaluating excited-state forces and NACVs without incurring the prohibitive cost of ab initio methods.
Recently, machine learning (ML) holds the potential to fundamentally reshape the landscape of NAMD simulations \cite{liu_excited-state_2025, shakiba_machine-learned_2024, westermayr_combining_2020, dral_mlatom_2024, mausenberger_spainn_2024, li_automatic_2021,wang_all-atom_2022}. 
However, previous efforts in solid-state ML-NAMD \cite{liu_breaking_2024}, including our own recent work \cite{zhang_advancing_2025}, have primarily focused on accelerating the CPA workflow. While separate progresses have been made in learning excited-state potentials \cite{Martyka_OMNI-P2x_2025, barrett_transferable_2025} and NACs for molecular systems \cite{Martinka_descriptor_2025, li_machine_2024, axelrod_excited_2022}, a unified framework capable of lifting the restrictions of the CPA in solids has yet to be realized.

In this Letter, we present on-the-fly $\mathrm{N^2AMD}$, a general machine learning framework that resolves the longstanding cost-accuracy dilemma in solid-state NAMD. Building upon an E(3)-equivariant neural network representation of the electronic Hamiltonian, our approach provides direct access to the exact excited-state forces and NACVs required for trajectory propagation through the learned Hamiltonian and its derivatives. This fundamentally obviates the CPA, making true on-the-fly NAMD in periodic solids a reality. We demonstrate that this framework achieves ab initio accuracy across diverse semiconductors at a fraction of the computational cost, opening the vast, uncharted landscape of strong electron-phonon coupling and non-equilibrium phase engineering to predictive theoretical exploration.


\begin{figure}[ht]
\includegraphics[width=1.0\linewidth]{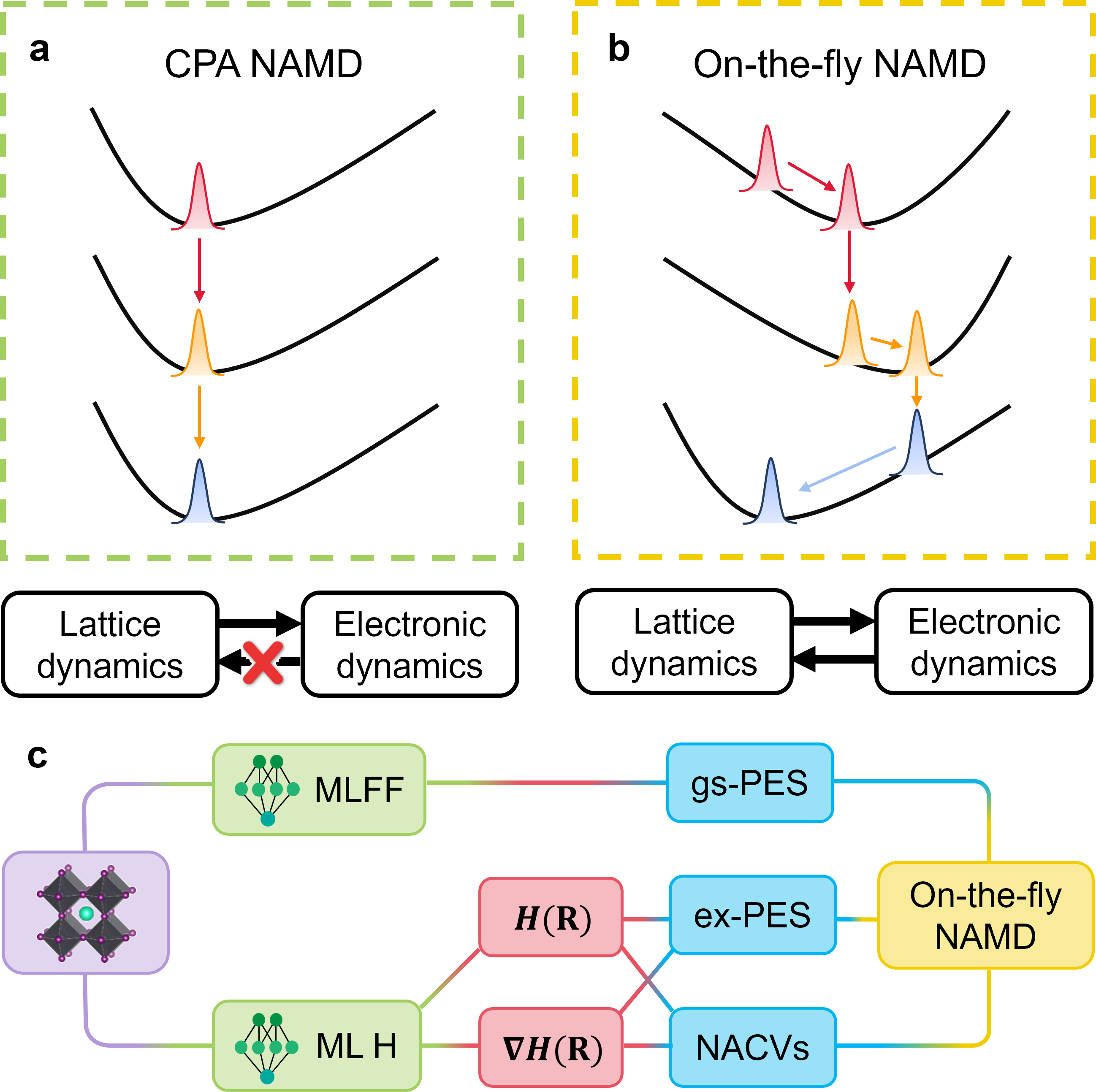}
\caption{\label{fig1} (a) Conventional NAMD (CPA) in solids restricts nuclear motion to a static PES. (b) On-the-fly NAMD allows nuclei to evolve on excited-state (ex) PESs dictated by electronic transitions, closing the loop of coupled electron-nuclear interactions. (c) The on-the-fly $\mathrm{N^2AMD}$ workflow: An equivariant neural network predicts the electronic Hamiltonian, with nuclear derivatives obtained via automatic differentiation. Subsequently, ex-PESs and NACVs are derived from these outputs (Eqs.~\ref{eq2} and \ref{eq4}) and combined with the MLFF-predicted gs-PES to drive the full NAMD simulation.
}
\end{figure}
\begin{figure*}[t]
\includegraphics[width=0.9\linewidth]{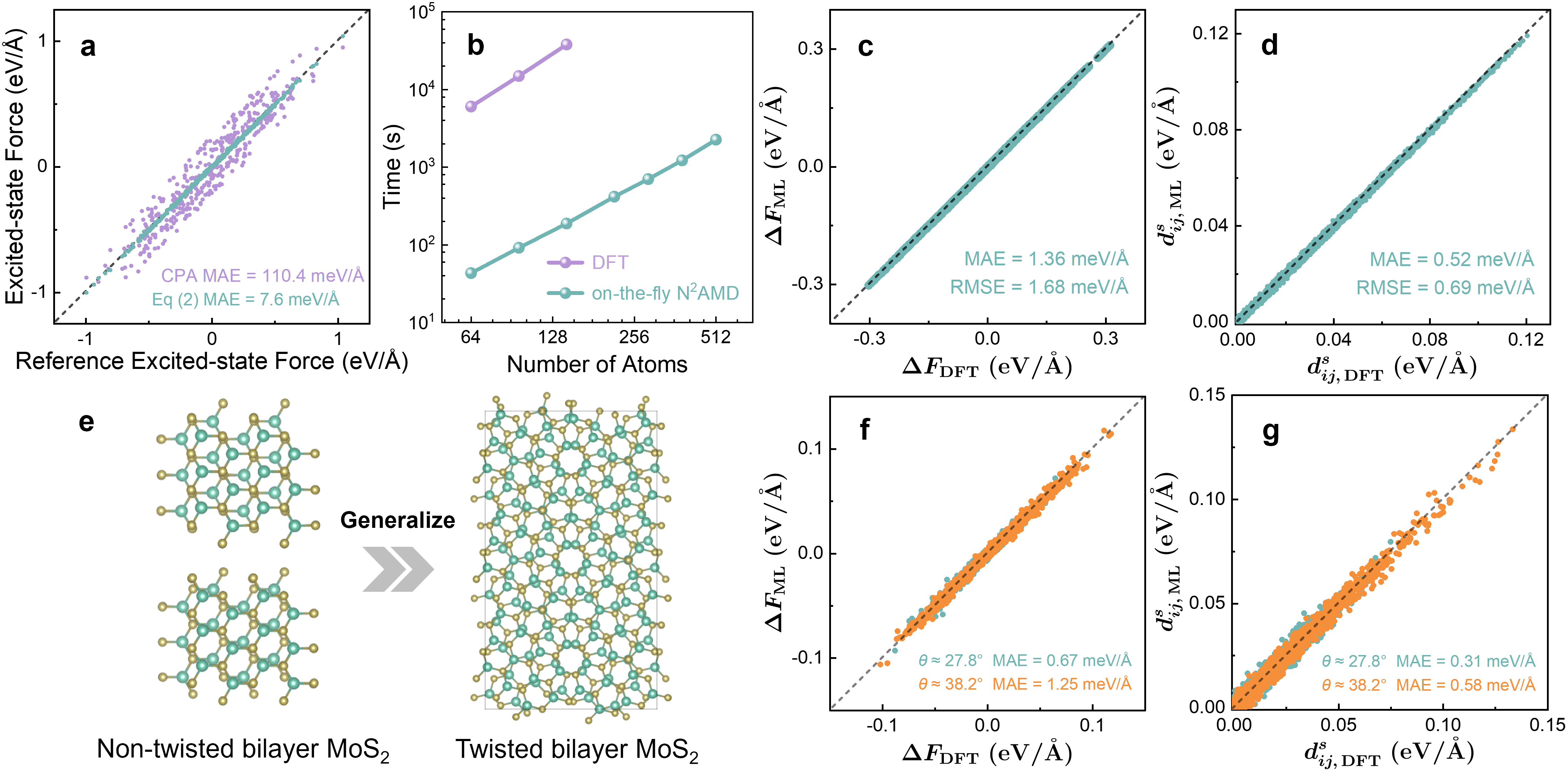}
\caption{\label{fig2} Accuracy, efficiency and transferability of on-the-fly $\mathrm{N^2AMD}$.
(a)-(d) Benchmarks on bulk silicon. (a) A comparison of reference $\Delta$-SCF excited-state forces against $\mathrm{N^2AMD}$ predictions (Eq.~\ref{eq2}, cyan) and the ground-state approximation (purple). 
(b) Computational time per ionic step for different sized systems.
(c) Prediction accuracy for the excitation force contribution $\Delta \mathbf{F}=\mathbf{F}_{ex}-\mathbf{F}_{gs}$, and (d) for the magnitudes of smoothed NACVs $d_{ij}^s$.
(e)-(g) Transferability demonstration on bilayer $\mathrm{MoS_2}$. $\mathrm{N^2AMD}$ is trained exclusively on nontwisted configurations and evaluated on unseen twisted structures with varying angles $\theta$.
}
\end{figure*}

In NAMD simulations within the framework of Density Functional Theory (DFT) \cite{craig_trajectory_2005, akimov_pyxaid_2013}, the time-dependent electron wavefunction is expanded in the basis of instantaneous adiabatic Kohn-Sham (KS) orbitals $\Psi(\mathbf{r}, \mathbf{R}, t) = \sum_nC_n(t)\psi_n(\mathbf{r}, \mathbf{R}(t))$. The wavefunction evolves according to the time-dependent Schrödinger equation: 
\begin{equation} 
    i\hbar \dot{C}_m(t) = \sum_n C_n(t) \left( \epsilon_m \delta_{mn} -i\hbar \mathbf{d}_{mn} \cdot \dot{\mathbf{R}}
    \right)
\label{eq1}
\end{equation}   
where $\epsilon_n = \langle \psi_n|\hat{H}|\psi_n\rangle$ is the $n$th KS eigen energy and  $\mathbf{d}_{mn} = \langle \psi_m | \nabla_{\mathbf{R}} | \psi_n \rangle$ represents the NACV between states $m$ and $n$. 
The coupled nuclear dynamics are governed by Newton's equations of motion: $\mathbf{F} = M\ddot{\mathbf{R}}$. 
Under the standard CPA, the back-reaction of the excited charge carriers on the lattice is neglected, restricting the nuclear forces to those of the ground state $\mathbf{F} \approx \mathbf{F}_{gs}$ (Fig.~\ref{fig1}a). To overcome this limitation and accurately describe the coupled dynamics of both subsystems, we compute the excited-state PESs using the delta self consistent field ($\Delta$-SCF) approach \cite{Gavnholt_delta_2008}. This method has demonstrated significant success in both condensed matter physics and quantum chemistry as a robust tool for investigating electronic excited states \cite{yang_foundation_2024, gao_tuning_2021, Vandaele_scf_2022}. Within the one-body band-excitation picture, for an excited state involving transitions from valence bands $\{i\}$ to conduction bands $\{a\}$ with occupation numbers $\{f_i\}$ and $\{f_a\}$, respectively, the force is given by
\begin{equation}
    \mathbf{F} \approx \mathbf{F}_{\mathrm{gs}}
    + \sum_{i\in\mathrm{VB}} (1-f_i)\nabla_{\mathbf{R}}\epsilon_i
    - \sum_{a\in\mathrm{CB}} f_a\nabla_{\mathbf{R}}\epsilon_a 
\label{eq2}
\end{equation}
Here, the total force is expressed as the sum of the ground-state force and the difference in KS orbital gradients. The double-counting energies vanish upon differentiation, as they do not explicitly depend on atomic positions. The formalism effectively decouples the ground-state and excitation energy scales, not only allowing for a direct assessment of the dynamical effects neglected by the CPA but also providing an efficient and robust framework for integration with neural networks. The primary approximation in this approach is the neglect of the self-consistent relaxation of the electronic Hamiltonian in response to the excited charge density. Nevertheless it remains valid for most non-strongly correlated solids (as demonstrated in the \textit{Benchmarks} section).
By expanding the adiabatic KS orbitals in a numerical atomic orbital (NAO) basis $\{\phi_\mu\}$ with coefficients $\{c_{i\mu}\}$, the orbital gradients in Eq.~\ref{eq2} are evaluated as (Detailed derivation in the Supplementary Material (SM) \cite{sm} Sec. 1):
\begin{equation} 
    -\nabla_{\mathbf{R}} \epsilon_i = 
    \sum_{\mu\nu} c_{i\mu}^* c_{i\nu} \times \left[-\nabla_{\mathbf{R}} H_{\mu\nu} + \epsilon_i \nabla_{\mathbf{R}} S_{\mu\nu} \right]
\label{eq3}
\end{equation} 
where $H_{\mu\nu}=\langle\phi_\mu|\hat{H}|\phi_{\nu}\rangle, S_{\mu\nu}=\langle\phi_\mu|\phi_{\nu}\rangle$ are Hamiltonian and overlap matrices.
Note that explicit k-point indices are omitted in this formalism because, in the current implementation, our NAMD simulations employ large supercells, allowing electronic structure calculations to be safely restricted to a single k-point (the $\Gamma$-point).
Beyond the neglect of excited-state forces, the CPA also fails to account for lattice heating resulting from carrier transitions, as it typically relies on a simple Boltzmann scaling of transition probabilities to mimic thermal equilibrium \cite{zheng_ab_2019}. In contrast, on-the-fly NAMD explicitly conserves total energy during transitions by rescaling the nuclear velocities at each hopping event: $\dot{\mathbf{R}}' = \dot{\mathbf{R}} - (\sigma_{mn}/M)\mathbf{d}_{mn}$. Such adjustment is applied along the direction of the NACV, which can be obtain by \cite{abad_calculation_2013} (Detailed derivation in SM \cite{sm} Sec. 1):
\begin{eqnarray} 
\label{eq4}
    \mathbf{d}_{mn} =&& \left\langle \psi_m | \nabla_{\mathbf{R}} | \psi_n \right\rangle 
    = - \frac{1}{\epsilon_m - \epsilon_n} \left\langle \psi_m \middle| \nabla_{\mathbf{R}} \hat{H} \middle| \psi_n \right\rangle \nonumber\\
    =&& \frac{1}{\epsilon_m - \epsilon_n} \sum_{\mu\nu} c_{m\mu}^* c_{n\nu} 
    \times \left[-\nabla_{\mathbf{R}} H_{\mu\nu}\right. \\
    &&\left.+ \epsilon_n \langle \nabla_{\mathbf{R}} \phi_{\mu}| \phi_{\nu} \rangle + \epsilon_m \langle \phi_{\mu}| \nabla_{\mathbf{R}} \phi_{\nu} \rangle 
    \right], \, m\ne n \nonumber
\end{eqnarray} 


Implementing NAMD beyond the CPA requires the repetitive evaluation of Hamiltonian gradients for excited-state forces and NACVs, a task that becomes computationally prohibitive via density functional perturbation theory (DFPT). 
Our on-the-fly $\mathrm{N^2AMD}$ framework circumvents this bottleneck by leveraging E(3)-equivariant neural networks (Fig.~\ref{fig1}c). 
While we retain the foundational ML building blocks, including the HamGNN electronic Hamiltonian \cite{zhong_transferable_2023} and a dedicated ground-state ML interatomic potential Allegro \cite{musaelian_learning_2023}, from our previous work \cite{zhang_advancing_2025}, the present work orchestrates them into a fundamentally different on-the-fly methodology. (See detailed comparison between two workflows in SM \cite{sm}, Sec. 11) As the first step of this workflow,
we employ HamGNN to construct deep neural Hamiltonians directly from instantaneous atomic configurations, bypassing iterative DFT cycles while rigorously preserving geometric symmetries. Upon diagonalizing the predicted Hamiltonian matrices to recover the KS energies \(\epsilon_i\) and orbitals \(\psi_i\), the workflow proceeds to evaluate the matrix gradients required for force and coupling calculations (Eqs.~\ref{eq2} and \ref{eq4}). These are efficiently obtained by computing the Hamiltonian gradient \(\nabla_{\mathbf{R}} H_{\mu\nu}\) via forward-mode automatic differentiation, alongside the overlap matrix gradient \(\langle \nabla_{\mathbf{R}} \phi_{\mu}| \phi_{\nu} \rangle\) and \(\langle \phi_{\mu}| \nabla_{\mathbf{R}} \phi_{\nu}\rangle\) using a two-center integral technique. Finally, efficient NAMD can be executed by combing these key quantities with the ground-state PES predicted by Allegro. Crucially, this hybrid "delta-prediction" architecture decouples the extensive ground-state PES from the intensive excitation contributions derived from the neural Hamiltonian. This separation resolves the fundamental scaling conflict inherent to extended systems, ensuring physical consistency and transferability while maximizing computational efficiency (See SM \cite{sm} Sec.~2 for workflow details).


\textit{Benchmarks.} We first validate the excited-state force formalism (Eq.~\ref{eq2}) using bulk silicon, a prototypical semiconductor. To simulate a non-equilibrium regime with non-zero forces, atoms in a 64-atom supercell are randomly displaced from ideal lattice positions with 0.02Å. As shown in Fig.~\ref{fig2}a, promoting one single electron from the valence band maximum (VBM) to the conduction band minimum (CBM) induces a significant reshaping of the PES. Our formalism captures this shift with exceptional fidelity, agreeing with rigorous, self-consistent $\Delta$-SCF benchmarks to within a mean absolute error (MAE) of 7.6 meV/Å. In contrast, the ground-state approximation (CPA) yields a discrepancy of 110.4 meV/Å, confirming that the CPA introduces substantial errors under relatively strong photoexcition (Here 0.39\% of the valence electrons excited).
Computational efficiency is benchmarked in Fig.~\ref{fig2}b. For a 96-atom silicon system, on-the-fly $\mathrm{N^2AMD}$ reduces the wall time per ionic step on one Intel(R) Xeon(R) Max 9462 @ 2.7GHz CPU from 4.14 hours by DFT to 91.43 seconds, achieving a speedup of 163$\times$. This advantage grows with system size, as our framework exhibits $\mathcal{O}(N^2)$ scaling, whereas DFT follows a much steeper trajectory. It is worth noting that the observed scaling is determined by our current implementation, which employs the full Hamiltonian matrix and its derivatives. Future algorithmic refinements leveraging the physical principle of "nearsightedness", specifically by exploiting matrix sparsity and symmetry, could significantly reduce this scaling for large-scale simulations (See SM \cite{sm} Sec.~3 for a detailed comparison of the scaling behaviors).

We next assess the accuracy of the on-the-fly $\mathrm{N^2AMD}$ framework in predicting essential NAMD components. The deep neural Hamiltonian reproduces DFT-computed real-space matrices with an MAE of 0.045 meV and yields a closely matched band structure (SM \cite{sm}, Figs.~S2a-b). Building on this foundation, we evaluate the critical ingredients for NAMD beyond the CPA: excited-state forces and NACVs.
For seven states around the Fermi surface, on-the-fly $\mathrm{N^2AMD}$ accurately captures the differences between ground- and excited-state forces $\Delta \mathbf{F}=\mathbf{F}_{ex}-\mathbf{F}_{gs}$, yielding an MAE of 1.36 meV/Å (Fig.~\ref{fig2}c). As for NACVs, we target the magnitude of smoothed NACVs defined as $\mathbf{d}_{ij}^{s} = |(\epsilon_i-\epsilon_j) \mathbf{d}_{ij}|$ to migrate to issues associated with its numerical instabilities near accidental energy degeneracies and the arbitrary phases of KS wavefunctions. It is found that our framework is capable of predicting couplings between selected state pairs with extremely high fidelity (Fig.~\ref{fig2}d, MAE: 0.52 meV/Å).

We further compare our on-the-fly $\mathrm{N^2AMD}$ framework against standard CPA NAMD by investigating hole relaxation dynamics in a type-II $\mathrm{MoS_2/WS_2}$ heterostructure (SM \cite{sm}, Fig.~S4b). Although both methods identify the same relaxation pathway ($\mathrm{MoS_2@K} \rightarrow \mathrm{WS_2@K} \rightarrow \mathrm{WS_2@\Gamma}$), significant discrepancies emerge in the kinetics (SM \cite{sm}, Figs.~S4c-f). CPA NAMD suggests carrier trapping at the intermediate $\mathrm{WS_2@K}$ state (SM \cite{sm}, Fig.~S4c), resulting in a dramatic underestimation of the relaxation rate. In contrast, on-the-fly $\mathrm{N^2AMD}$ correctly captures the rapid decay to the final $\mathrm{WS_2@\Gamma}$. This improvement arises from the complete treatment of electron-phonon coupling, which captures the generation of non-equilibrium phonon populations that drive efficient energy dissipation. To ensure a rigorous comparison, both methods employ NAC phase corrections \cite{akimov_simple_2018}, preventing the artificial acceleration noted in earlier work \cite{zheng_phonon-assisted_2017}. The superior performance of $\mathrm{N^2AMD}$ is thus strictly attributable to the inclusion of explicit excited-state forces.

Finally, we demonstrate the framework's transferability across diverse solid state systems. Two challenging scenarios were tested: (i) Moiré physics in bilayer $\mathrm{MoS_2}$ (Fig.~\ref{fig2}e-g), where a model trained exclusively on non-twisted bilayers successfully generalizes to twisted structures; and (ii) configurational disorder and variable stoichiometry (the Ga/Al mixing ratio $x$) in the solid solution $\mathrm{Ga}_x\mathrm{Al}_{1-x}\mathrm{As}$ (SM \cite{sm}, Figs.~S3e-g), where a model trained on configurations with $x=0$, $1/4$, $2/4$, $3/4$ and $1$ accurately predicts unseen compositions ($x=3/8$, $5/8$). In both cases, excited-state forces and NACVs are reproduced with meV/Å- or sub-meV/Å-level accuracy, ensuring robust simulation of far-from-equilibrium dynamics in novel materials.


\begin{figure}[ht]
\includegraphics[width=1.0\linewidth]{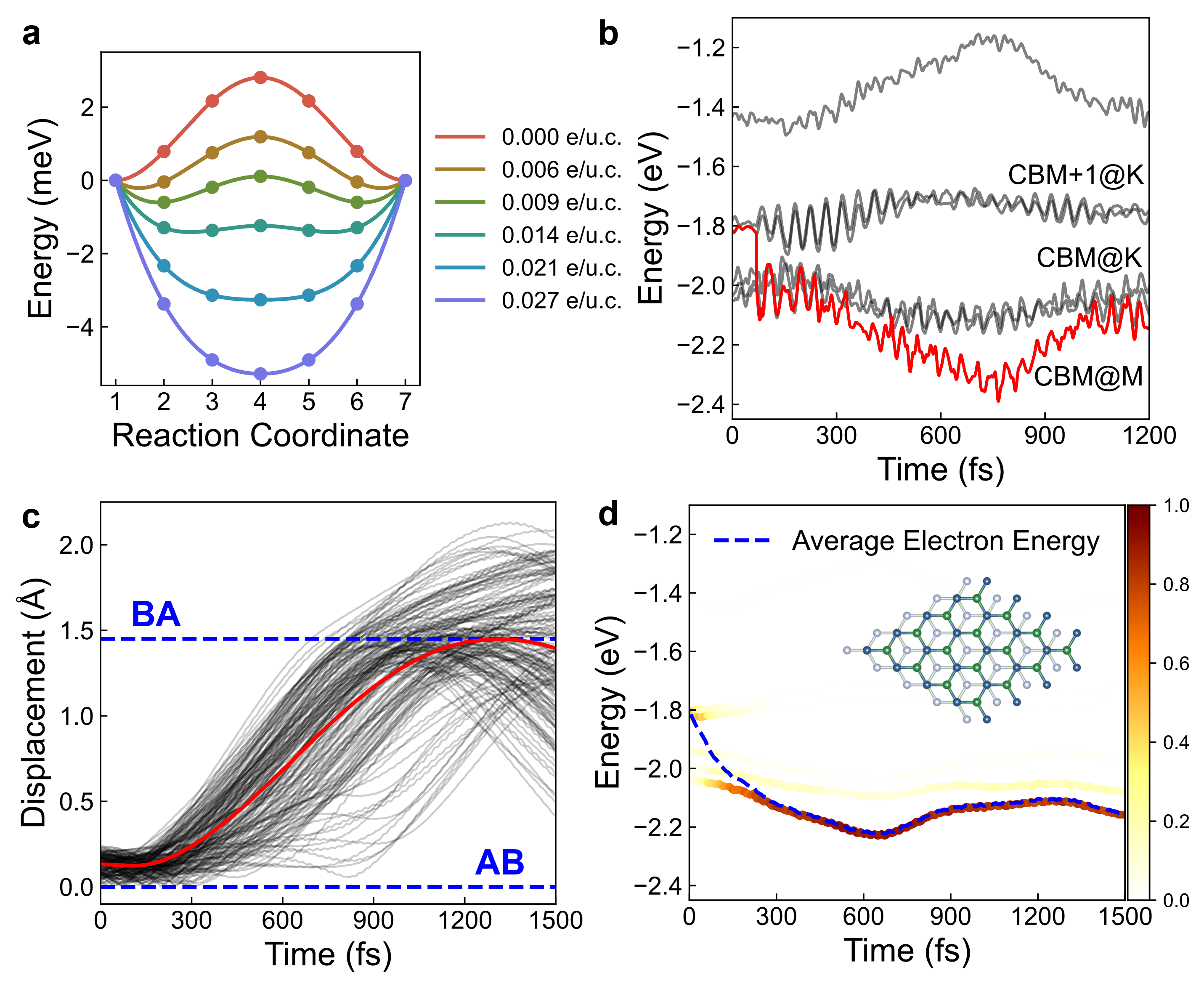}
\caption{\label{fig3} Photo-induced ferroelectric switching in bilayer $h$-BN.  (a) Potential energy curves along the AB-BA transition pathway under varying photoexcitation densities (0.000-0.027 e/u.c.). Energies of the equilibrium AB and BA states are set at zero. 
(b)-(d) On-the-fly $\mathrm{N^2AMD}$ results: (b) KS eigenvalue evolution in one representative trajectory with electron hopping (red curve), (c) in-plane sliding displacement progressions across the trajectory ensemble relative to equilibrium AB and BA states (blue dashed lines), ensemble average displacement (red line) and (d) average electron energy evolution, with the color scale indicating the per-state excited electron population.
}
\end{figure}

\textit{Application 1: Photo-induced ferroelectric switching.}
We next apply on-the-fly $\mathrm{N^2AMD}$ to a regime fundamentally inaccessible to the conventional CPA NAMD in solids: photo-induced phase transitions driven by the reshaping of excited-state potential energy landscapes. A prime example of this regime is sliding ferroelectricity, a promising route for next-generation nonvolatile memory \cite{bian_developing_2024} where ultrafast manipulation via laser pulses requires accurate theoretical simulation \cite{sie_ultrafast_2019, gao_large_2024, yang_light-Induced_2024, tao_tunable_2024, yao_Reprogrammable_2025}. We investigate photo-induced ferroelectric switching in bilayer hexagonal boron nitride ($h$-BN), focusing on how photoexcitation tunes the energy barrier between AB and BA stacking orders. In our periodic models, the photoexcitation is appropriately quantified by the excitation density. Static calculations show that an excitation density of 0.027 e/u.c. eliminates the barrier entirely, creating a driving force for spontaneous switching (Fig.~\ref{fig3}a).
Simulations at this excitation concentration, 100 K reveal a stark dichotomy between methods. CPA NAMD predicts zero switching events within 1.5 ps (SM \cite{sm}, Fig.~S5b). Conversely, on-the-fly $\mathrm{N^2AMD}$ suggests that 82\% of trajectories undergo a complete polarization reversal (in-plane displacement reaches $\xi=1.45$ Å) in the same timeframe (Fig.~\ref{fig3}c). 
Analysis of a representative trajectory reveals that the CBM@M KS eigenvalue drops significantly between 300-800 fs before recovering, providing the requisite force for lattice sliding (Fig.~\ref{fig3}b). In contrast, CPA NAMD eigenvalues merely oscillate around equilibrium (SM \cite{sm}, Fig.~S5a). Notably, both methods capture an initial rapid (200 fs) electron relaxation from CBM+1@K to CBM@M (Fig.~\ref{fig3}d and SM \cite{sm}, Fig.~S5c), which explains the brief displacement plateau preceding sliding. To further validate the necessity of our approach, one constrained Born-Oppenheimer (BO) MD simulation is also performed (SM \cite{sm}, Figs.~S5d-f). While it correctly predicts the sliding, it inherently fails to describe the critical hot-carrier cooling observed in NAMD results. Thus, on-the-fly $\mathrm{N^2AMD}$ resolves the complete mechanism surpassing both CPA NAMD and constrained BOMD.


\begin{figure}[ht]
\includegraphics[width=0.9\linewidth]{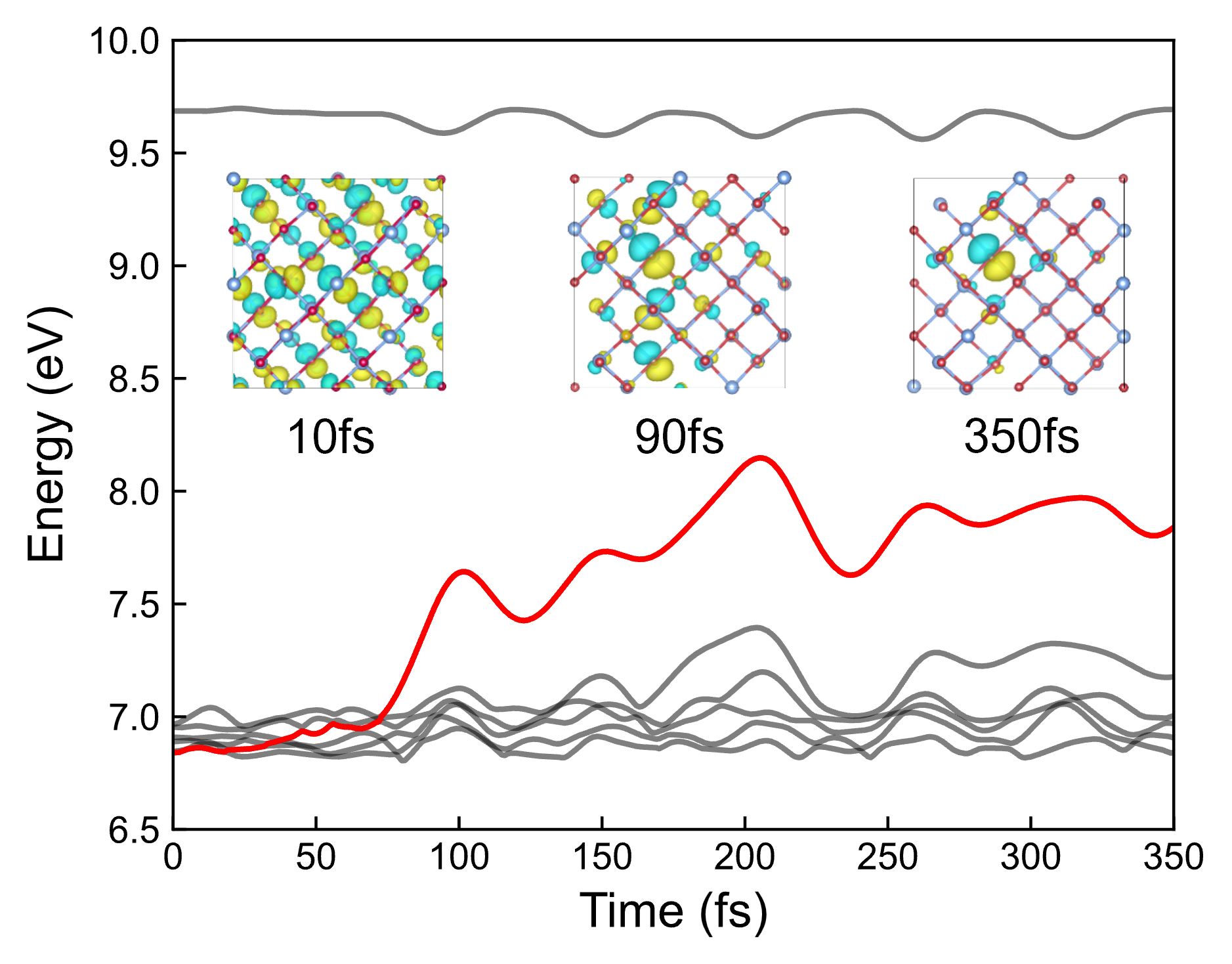}
\caption{\label{fig4} Real-time evolution of carrier localization in anatase $\mathrm{TiO_2}$ demonstrating ultrafast polaron formation. Lines from bottom to top: states from VBM-5 to VBM, and the CBM. Insets: hole-hosting state wavefunctions at 10, 90 and 350 fs. 
}
\end{figure}

\textit{Application 2: Quasiparticle formation dynamics.}
Finally, we demonstrate the capacity of on-the-fly $\mathrm{N^2AMD}$ to capture quasiparticle dynamics at the hybrid functional level. Although the extreme computational cost of hybrid functionals generally limits conventional NAMD and Ehrenfest dynamics of extended solids to the Perdew-Burke-Ernzerhof (PBE) functional \cite{perdew_generalized_1996}, our framework leverages deep neural Hamiltonians to incorporate hybrid functional accuracy without the associated computational bottleneck. We apply this approach to investigate the formation of small hole polarons in anatase $\mathrm{TiO_2}$, a process where mitigating self-interaction errors via hybrid functionals is critical, yet historically intractable for NAMD \cite{Franchini_polarons_2021, elmaslmane_first-principles_2018}.
Static Heyd-Scuseria-Ernzerhof (HSE) functional \cite{heyd_hybrid_2003} calculations identify a polaron state 0.77 eV above the VBM, characterized by a $p$-orbital shaped wavefunction centered on an oxygen atom with 0.14 Å lattice distortions (SM \cite{sm}, Figs.~S6a-b). 
Our simulation provides a time-resolved view of its formation. Following photoexcitation, the hot hole relaxes to the VBM within 70 fs, then rapidly plunges into the band gap, entering the self-trapping state (Fig.~\ref{fig4}, red line). The hole wavefunction transitions from delocalized at 10 fs to highly localized at 90 fs, achieving a stable polaron configuration by 350 fs (Insets, Fig.~\ref{fig4}). By rendering such demanding simulations feasible, on-the-fly $\mathrm{N^2AMD}$ offers a powerful tool to explore complicated polaron-mediated dynamics.


In conclusion, on-the-fly $\mathrm{N^2AMD}$ resolves the long-standing computational bottleneck prohibiting exact non-adiabatic molecular dynamics in extended systems. By mapping the \textit{ab initio} electronic Hamiltonian to an E(3)-equivariant neural network, we restore the essential feedback loop between excited carriers and the lattice without the prohibitive scaling of density functional theory. The method’s versatility is evidenced by its successful application to distinct non-equilibrium regimes, including correcting hot carrier relaxation rates in heterostructures, driving non-thermal phase transitions in ferroelectrics, and capturing self-trapping polaron dynamics in oxides. This framework bridges the gap between static electronic structure and non-equilibrium dynamics, paving the way for the first-principles design and discovery of materials operating far from equilibrium.


\begin{acknowledgments}
We thank Dr. Daye Zheng for implementing overlap gradient outputs in ABACUS. We acknowledge support from the National Natural Science Foundation of China (12188101; 12274081; 11991061; 22333003, 22361132528; 12271109, 12526211); Shanghai Science and Technology Program (23JC1400900); National Key Research and Development Program of China (2024YFA1409800); Shanghai Pilot Program for Basic Research - Fudan University 21TQ1400100 (No. 22TQ017); Scientific Research Innovation Capability Support Project for Young Faculty (SRICSPYF-ZY2025159) and Xuemin Institute of Advanced Studies, Fudan University. O.V.P acknowledges the support from the US National Science Foundation (CHE-2603101). X. G. G. acknowledges support of the Innovation Program for Quantum Science and Technology (No. 2024ZD0300100).
\end{acknowledgments}

\textit{Data availability}—The source code of on-the-fly $\mathrm{N^2AMD}$, the raw data, and an example case with detailed instructions on how to use our code are available at Zenodo \cite{zenodo}.

\appendix



%

\end{document}